\documentstyle[emulateapj,psfig,apjfonts]{article}
\submitted{Received 2006 August 4; Accepted: 2006 November 16}
\input{psfig.sty}

\def\gs{\mathrel{\raise0.35ex\hbox{$\scriptstyle >$}\kern-0.6em
\lower0.40ex\hbox{{$\scriptstyle \sim$}}}}
\def\ls{\mathrel{\raise0.35ex\hbox{$\scriptstyle <$}\kern-0.6em
\lower0.40ex\hbox{{$\scriptstyle \sim$}}}}

\makeatother

\lefthead{Smail et al.}
\righthead{The ``Cosmic Eye''}

\begin{document}

\title{A very bright, highly magnified Lyman-break galaxy at $z=3.07$.}

\author{
Ian Smail,\altaffilmark{1} A.\,M.\ Swinbank,\altaffilmark{1}
J.\ Richard,\altaffilmark{2} H.\ Ebeling,\altaffilmark{3}
J.-P.\ Kneib,\altaffilmark{4} A.\,C.\ Edge,\altaffilmark{1}
D.\ Stark,\altaffilmark{2}  R.\,S.\ Ellis,\altaffilmark{2}
S.\ Dye,\altaffilmark{5}
G.\,P.\ Smith\altaffilmark{6} \& C.\ Mullis.\altaffilmark{7}
}

\altaffiltext{1}{Institute for Computational Cosmology, Durham University, South Road,
        Durham, DH1 3LE, UK}
\altaffiltext{2}{California Institute of
Technology, Department of Astronomy, MC 105-24, Pasadena, CA 91125}
\altaffiltext{3}{Institute for Astronomy, 2680 Woodlawn Drive, Honolulu, HI 96822}
\altaffiltext{4}{Laboratoire d'Astrophysique de Marseille, Traverse du
        Siphon -- B.P.8 13376, Marseille Cedex 12, France}
\altaffiltext{5}{Cardiff University, School of Physics \& Astronomy, Queens Buildings,
        The Parade, Cardiff, CF24 3AA, U.K.}
\altaffiltext{6}{School of Physics and Astronomy, University of
        Birmingham, Edgbaston, Birmingham, B15 2TT, U.K.}
\altaffiltext{7}{Department of Astronomy, University of Michigan, 918
        Dennison, 500 Church Street, Ann Arbor, MI 48109-1042}

\setcounter{footnote}{5}

\begin{abstract}
We report the discovery using {\it Hubble Space Telescope} imaging and
Keck spectroscopy of a very bright, highly magnified ($\sim 30\times$)
Lyman Break Galaxy (LBG) at $z=3.07$ in the field of the massive
$z=0.33$ cluster MACS\,J2135.2$-$0102.  The system comprises two
high-surface brightness arcs with a maximum extent of $3''$, bracketing
a central object which we identify as a massive early-type galaxy at
$z=0.73$.  We construct a lens model which reproduces
the main features of the system using a combination of a galaxy-scale
lens and the foreground cluster.  We show that the morphological,
spectral and photometric properties of the arcs are consistent with
them arising from the lensing of a single $\sim L_V^\ast$ LBG.  The most
important feature of this system is that the lensing magnification
results in an apparent magnitude of $r=20.3$, making this one of the
brightest LBGs known.  Such a high magnification provides the
opportunity of obtaining very high signal to noise (and potentially
spatially resolved) spectroscopy of a high redshift galaxy to study its
physical properties.  We present initial imaging and spectroscopy
demonstrating the basic properties of the system and discuss the
opportunities for future observations.
\end{abstract}

\keywords{cosmology: observations --- galaxies: individual (LBG\,J213512.73$-$010143) ---
          galaxies: evolution --- galaxies: formation}
 
%
%
%
\section{Introduction}

Some of the most compelling science drivers from the extragalactic
research field for the construction of Extremely Large Telescopes (ELT)
involve investigating the internal dynamics of the gas and stars, and
the chemical abundances of these components, in faint high-redshift
galaxies (Hook 2005).  This requires a combination of high
spatial-resolution and high signal-to-noise spectroscopy which only the
largest telescopes can provide.  These observations will yield
insights into the physical properties of star-formation in
the early Universe.  Yet some of these scientific questions can be
tackled with the current generation of 10-m-class telescopes, when
aided by the natural magnification provided by a gravitational lens.
Such lens-aided studies are already providing some of our most
sensitive views of the distant Universe, with lensing surveys for faint
sources out to the highest redshifts (Kneib et al.\ 2004b) and across a
range of wavebands (e.g.\ Ellis et
al.\ 2001; Smail et al.\ 2002; Metcalfe et al.\ 2003; Santos et al.\
2004; Kneib et al.\ 2004a).

In terms of our understanding of the internal properties of high
redshift galaxies, the studies of the gravitationally lensed Lyman
break galaxy (LBG) cB58 by Pettini et al.\ (2000, 2002) have
demonstrated that high-signal-to-noise spectra of typical LBGs can be
obtained with 10-m telescopes, if they are highly magnified as a result
of being serendipitously positioned behind a suitable foreground
gravitational lens.  cB58 has an apparent magnitude of $r=20.4$ and
represents a $\sim (30\pm 10) \times$ magnified image of an $L^\ast$
LBG at $z=2.72$.  The studies of this object have yielded a wealth of
information on the metallicity and energetics of the interstellar
medium in a young star-forming galaxy (Pettini et al.\ 2002).  
The only drawback with these studies is the difficulty in
drawing wide-ranging conclusions from a single object, and hence the
urgent need to find other examples of similarly highly magnified
galaxies.  To this end, several deep imaging surveys of clusters have
been undertaken to search for highly magnified LBGs using $U$- and
$B$-band dropout selection (e.g.\ Smail, Edge \& Ellis 1998; Stern et
al.\ 2004).  The difficulty for these surveys is the presence of large
numbers of red galaxies in the clusters, whose spectral properties are
sufficiently similar to the target population to be a significant
source of contamination and so far they have not yet yielded any highly
magnified, high-redshift galaxies as bright as cB58.  Similarly,
attempts to use the wide area coverage of the Sloan Digital Sky Survey
(SDSS) to identify rare intrinsically bright or highly magnified LBGs
(Bentz et al.\ 2004), have until recently only uncovered an unusual
class of AGN with weak restframe UV emission lines (Ivison et al.\
2005).  However, Allam et al.\ (2007) have just announced the discovery
of a bright lensed LBG using data from SDSS.

In this letter we present the discovery of a $z=3.07$ galaxy,
LBG\,J213512.73$-$010143, which appears as two highly-magnified arcs
with an apparent magnitude of $r=20.3$, making it slightly brighter
than cB58.  We assume a cosmology with $\Omega_m=0.27$,
$\Omega_\Lambda=0.73$ and $H_o=71$\,km\,s$^{-1}$\,Mpc$^{-1}$, giving
angular scales of 4.7, 7.3 and 7.8\,kpc arcsec$^{-1}$ at $z=0.33$,
$z=0.73$ and $z=3.08$ respectively.  All quoted magnitudes are 
AB.

\section{Observations and Reduction}

\subsection{{\it HST} Imaging}

We obtained a 1.2-ks {\it Hubble Space Telescope (HST)} ACS F606W image
of MACS\,J2135.2$-$0102 on 2006 May 8th, as part of our Snapshot
program (GO\# 10491, PI: H.\ Ebeling).  This program targets
high-luminosity X-ray clusters to both identify bright lensed galaxies
and constrain the cluster mass distributions.  MACS\,J2135.2$-$0102 is
a high-luminosity X-ray cluster at $z=0.325$ cataloged by Ebeling, Edge
\& Henry (2001).  The ACS exposure comprised three 400\,s exposures
dithered with a LINE pattern and a 3$''$ spacing.  The data were
reduced using {\sc multidrizzle} v2.7 to
provide an image with 0.05$''$ sampling and good cosmetic properties.

%
%
\centerline{\psfig{file=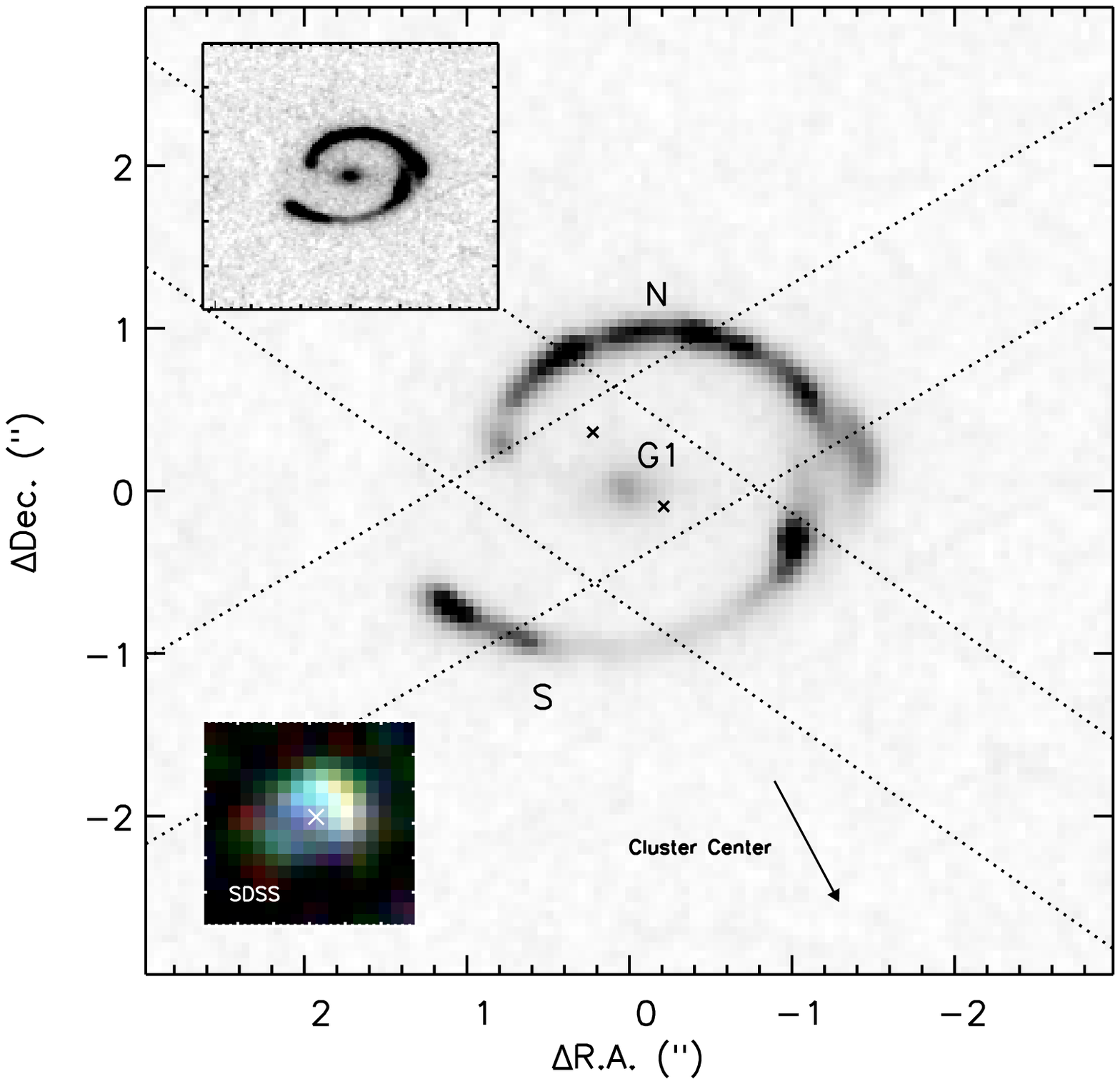,width=3.5in,angle=90}}
\noindent{\small\addtolength{\lineskip}{-2pt}
 {\sc Fig.~1.}--- 
An  {\it HST} ACS F606W image of
LBG\,J213512.73$-$010143 showing the twin arcs (labeled as North, N,
and South, S) around a compact source, G1.  Arc N has several bright
knots with a symmetric appearance, along with a low-surface brightness
extension to the west which has a different radius of curvature.  Arc S
has a bright knot at each end, connected by low-surface brightness
emission.  The two arcs are not concentric and we mark their centers of
curvature by ``X''.  The pairs of dotted lines indicate the alignment
of the spectroscopic slits in our Keck LRIS observations.  The inset to
the upper-left displays the same field at a higher contrast, while the
inset at the lower-left shows a true color SDSS $gri$ image of the
source and we indicate the direction to the cluster center.

}

A visual inspection of the {\it HST} image by
two of us (ACE/AMS) identified an unusual object at 21\,35\,12.730
$-$01\,01\,42.9 $\pm 0.5''$ (J2000), lying approximately 75$''$ due
north of the brightest cluster galaxy (21\,35\,12.08 $-$01\,02\,58,
J2000).  In the {\it HST} ACS image,
Fig.~1, the object appears as two non-concentric arcs, enclosing a
central compact object and is reminiscent of a ``Cosmic Eye''. We
denote the northern and southern arcs as N and S respectively and the
central source as G1.  Interestingly, the centers of curvature of N and
S do not coincide, lying WSW and NE of G1 respectively (Fig.~1).  The
radii of curvature are 1.0$''$ for N, which is nearly perfectly
circular, and 1.5$''$ for S, which is more flattened.  The faint, western
extension of N has a radius of curvature of 1.3$''$.

A search of the SDSS DR4 shows that this object is well-detected with a
total magnitude of $r=20.27$, consistent with the {\it HST} measurement
of $R_{606}=20.54\pm 0.02$ for the light from both arcs, and
$R_{606}=21.06\pm 0.05$, $R_{606}=21.60\pm 0.10$ and $R_{606}=22.34\pm
0.15$ for N, S and G1 respectively.  The arcs are unresolved across
their width, with apparent FWHM of $\sim 0.15''$ along their whole
lengths: 2.2$''$ for N and 2.8$''$ for S, yielding axial ratios of
$>15$ and $>19$ respectively.  In contrast, G1 has a FWHM of $\sim
0.25''$ with an ellipticity of $\epsilon \sim 0.3$ at a P.A.\ of
105\,deg.\ (i.e.\ aligned along the major axis of the arcs).

\setcounter{figure}{1}

%
%
\begin{figure*}[tbh]
\centerline{\psfig{file=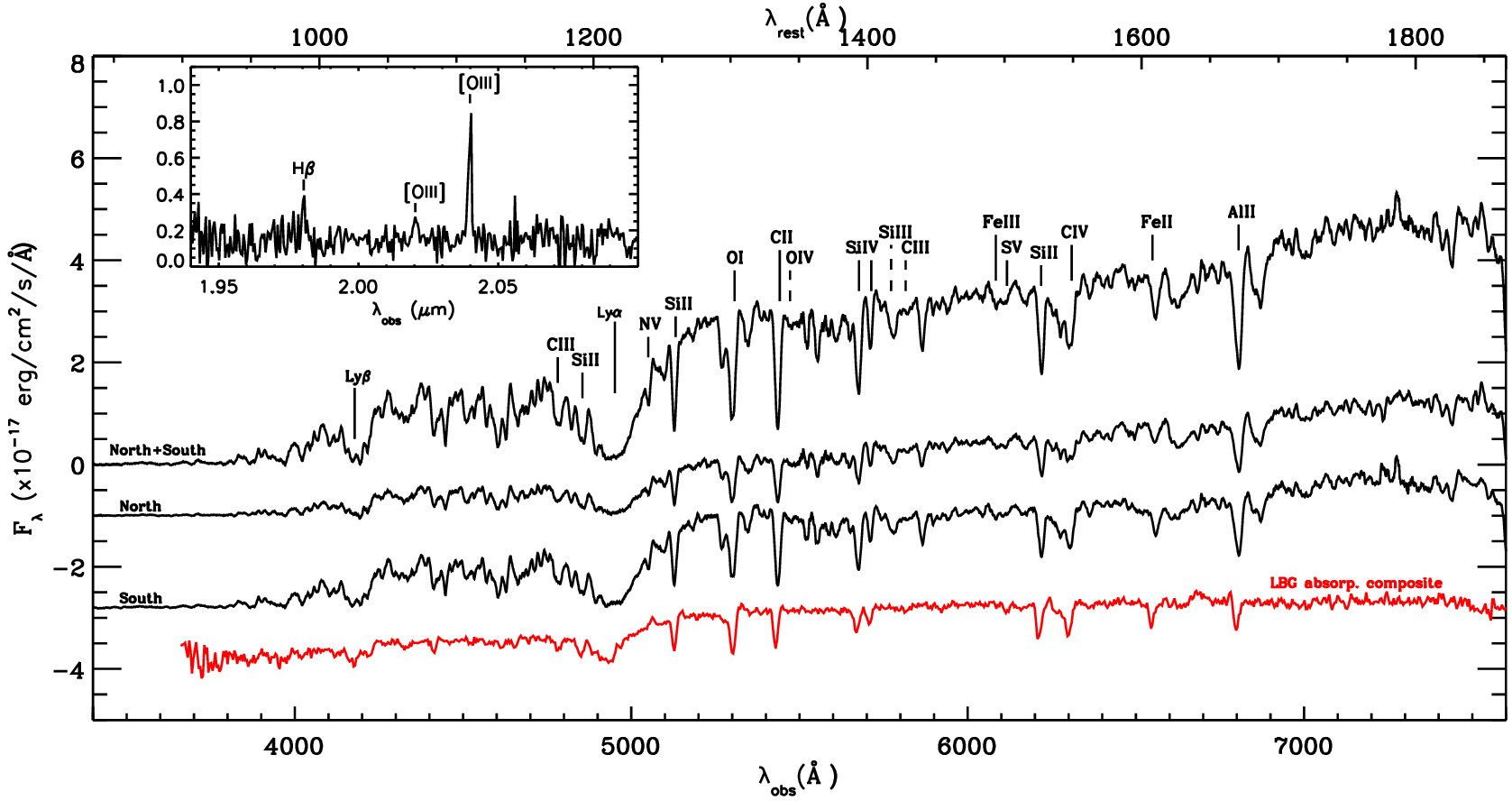,width=6.0in,angle=0}}
\caption{\small The combined Keck LRIS spectrum of the two arcs (top)
as well as the spectra for the individual arcs, N and S (center). We
mark the positions of the absorption features we identify in the
spectra, both the strong lines arising from the ISM in the galaxy and
the weaker photospheric features (marked with dotted lines).  The top
axis gives the restframe wavelength scale at the systemic redshift of
the source.  For comparison we show the composite spectrum of LBGs with
Ly$\alpha$ absorption from Shapley et al.\ (2003), note the broad
similarity in the main spectral features -- apart from the stronger
damping wings of the Ly$\alpha$ line in the lensed LBG.  We also show
in the inset the NIRSPEC spectrum of the eastern component of S,
identifying the H$\beta$ and [O{\sc iii}]4959,5007 lines which yield a
systemic redshift for the source of $z=3.0743$. The flux scale of the
inset is the same as the main figure. 
}
\end{figure*}

\subsection{Keck Spectroscopy and Imaging}

The morphology of this source is strongly suggestive of gravitational
lensing, with G1 identified as the primary lens.  In addition, the SDSS
$griz$ colors are blue ($(g-r)=1.20$, $(r-i)=0.53$ and $(i-z)=0.29$)
but with a red $(u-g)=2.30$ color indicating a spectral break
short-ward of the $g$-band and hence a probable high redshift, $z\sim
3$--4.  As such we observed the system as a high priority
with the LRIS spectrograph on the Keck I telescope on the night of 2006
June 30 in good transparency and $\ls 1.0''$ seeing.  The system was
observed at two position angles (P.A.\ of 55 and 120 degs, see Fig.~1)
to cover both arcs and G1. We employed the 600\,l/mm grism in the blue
arm, blazed at 4000\AA, and the 400\,l/mm grating in the red arm,
blazed at 8500\AA, giving wavelength coverage from the atmospheric
cutoff out to $\sim 9000$\AA. The total integration at each P.A.\ was
3.6\,ks. The spectra were reduced with standard {\sc iraf} and {\sc
idl} routines to yield fluxed, wavelength-calibrated spectra.  We show
the individual spectra for each arc and the total combined spectrum in
Fig.~2.  A spectrum was also extracted for G1.

In addition, we obtained a 2.4-ks exposure with NIRSPEC on Keck II on
the night of 2006 July 24.  This observation, taken at P.A.\ of 55
deg., covered the entire $K$-band window and detected continuum and a
series of narrow emission lines from the eastern component of S (see
the inset in Fig.~2).  Finally, we obtained a $K'$-band image of the
system with the NIRC near-infrared imager on Keck I on the night of
2006 July 4.  The total exposure time was 1.14\,ks as 19 individual
60-s exposures in 0.5$''$ seeing. The image detects and resolves both
arcs and the central galaxy, G1.

\section{Analysis and Discussion}

\subsection{Source Properties}

The LRIS spectra (Fig.~2) show that the two arcs have blue continua
with a series of strong absorption features and spectral breaks which
unambiguously identify the source redshift at $z\sim 3.074$.  
As Fig.~2 shows, the absorption features in the combined spectrum are
similar to those in the subset of the LBG population which exhibit
Ly$\alpha$ in absorption (Shapley et al.\ 2003).  In particular, we see
strong absorption from species in the interstellar medium (ISM) of the
galaxy, as well as broad, blueshifted absorption in the C{\sc iv}\,1550
and O{\sc i}\,1302 lines, both of which are observed in typical LBGs.
One distinctive difference is that the Ly$\alpha$ absorption line in
the arcs is stronger than in typical LBGs and indicates a significant
H{\sc i} column density of $\log($N(H{\sc i}$)) \sim 21.7$ ($\sim 7$
times higher than cB58, Pettini et al.\ 2000).  We qualitatively
compare the spectrum with the predictions from the {\sc Starburst99} model
(Leitherer et al.\ 1999, 2002) and the {\it IUE} spectra of O/B stars
in de Mello et al.\ (2000). The strength of the C{\sc iv}\,1550
absorption, the lack of an associated red emission wing and the
relative weakness of the Si{\sc iv}\,1400 absorption suggest that the
luminosity weighted stellar population in the arcs is dominated by
early B-type supergiants, indicating either a $\sim 10$\,Myr-old burst
of star formation or on-going activity over a similar or slightly
longer period, but with a stellar IMF deficient in O-type stars.

Analysis of the spectra for the individual subcomponents of the arcs
shows no evidence for velocity offsets between the internal components
of N and S, as expected if they are both multiply imaged.  We derive
redshifts, from fits to individual photospheric absorption lines (O{\sc
iv}\,1343.4, Si{\sc iii}\,1417.2, C{\sc iii}\,1427.8 and S\,{\sc
v}\,1501.8), for S of $z=3.0747\pm0.0005$ and for N of $z=
3.0744\pm0.0003$ (all errors are bootstrap estimates).  We caution that
these features are individually weak, but they suggest that there is no
significant velocity offset between the two arcs ($\ls
50$\,km\,s$^{-1}$). However, a comparison of the spectra of the two
arcs does show some marked differences: the profiles of the Ly$\alpha$,
Ly$\beta$ and the stronger ISM lines differ between N and S, and the
continuum appears redder in S, indicating that N and S are images of
different sources, either two galaxies, or more likely two regions
within a single galaxy (consistent with the lens reconstruction).

We also measure the redshifts of the strong ISM lines in the two arcs
obtaining $z=3.0727\pm 0.0004$ and $z=3.0726\pm 0.0003$ for single
Gaussian fits to the lines in S and N respectively, using Si{\sc
ii}\,1260.4, O{\sc i}\,1302.2 C{\sc ii}\,1334.5, Si{\sc iv}\,1402.8,
Si{\sc ii}\,1526.7 and Al{\sc ii}\,1670.8.  Looking at the stronger ISM
lines in the composite spectrum in more detail shows that they comprise
at least three separate components: a weak one at $z=3.0666\pm 0.0007$
and two stronger ones at $z=3.0715\pm 0.0010$ and $z=3.0752\pm 0.008$.
The reddest component has a redshift consistent with the photospheric
and nebular emission line estimates (see below) suggesting it arises in
gas at the systemic redshift of the system.  For the bluer ISM lines,
we derive blueshifts of $-200\pm 70$\,km\,s$^{-1}$ for the stronger
line and $-570\pm 60$\,km\,s$^{-1}$ for the weaker.  These are
consistent with outflows of material from the galaxy in the form of a
wind and the velocities are comparable to those seen in typical LBGs by
Steidel et al.\ (2007).  Turning to the continuum emission we measure a
restframe UV spectral slope for the composite spectrum between
1200--2000\AA\ of $\beta=-1.6\pm 0.1$ which implies A$_{1600}\sim 1.7$
or $E(B-V)\sim 0.4$ (Calzetti et al.\ 2000).  We also see a sharp
decline in flux as we move blue-ward of Ly$\beta$, with little
detectable emission shortward of 930\AA\ in the
restframe.

In the near-infrared spectrum available for the eastern component of S
we see three narrow emission lines corresponding to H$\beta$, [O{\sc
iii}]\,4959 and [O{\sc iii}]5007, with unresolved restframe FWHM of
$\ls 220$\,km\,s$^{-1}$ (corrected for the instrumental resolution).
These features indicate a redshift for the emission line
gas in the system of $z=3.0743\pm 0.0001$ which we adopt as the
systemic redshift for the galaxy.

Finally, the LRIS spectrum of G1 shows a strong continuum break around
6900\AA\ and several features which we identify
as Ca H\&K, G-band and [O{\sc ii}]\,3727 at
$z=0.7268\pm0.0007$.  The [O{\sc ii}]\,3727 line has an equivalent
width of just 7.5\AA, indicating modest star formation activity.  Thus
the likely lens appears to be an early type spiral behind the cluster,
and from the width of the G-band and Ca H\&K lines in the spectrum we
estimate a central velocity dispersion of $\sim 230\pm
30$\,km\,s$^{-1}$.  In addition to this galaxy, we also serendipitously
detect a further two galaxies which show [O{\sc ii}]\,3727 at $z\sim
0.73$ indicating the presence of a structure at this redshift.

Turning to the broadband imaging, we compare the morphology of the arcs
as seen in the NIRC $K'$-band image to the {\it HST} ACS F606W image.
This shows no evidence for strong color variations within or between
the two arcs, although the difference in resolution in the two bands
and the presence of the lens makes this comparison difficult.  In
addition, we see no evidence for a second lens in the system -- with
the position of G1 agreeing well between the optical and near-infrared
images.  We find that the arcs dominate the integrated light from the
system from the $g$- to $K$-bands: measuring $K'=18.9\pm 0.1$ and
$(R_{606}-K') \sim 1.7$ for the combined arcs and $K'=19.7\pm 0.1$ and
$(R_{606}-K') \sim 2.8$ for G1.  The color of G1 matches that of an
early type spiral or S0 at $z=0.73$, while the $K'$-band magnitude
yields $M_K\sim -23.9 + 5\log(h)$, indicating it is a $\sim L_K^\ast$
galaxy (consistent with its measured velocity dispersion and the
Faber-Jackson relation).

\subsection{Lensing Model}

To interpret the
properties of this system in more detail we have developed a lens model
for the system using the semi-linear inversion method of Warren \& Dye
(2003). The lens model includes not only a mass component for G1, but also a
significant external shear from the mass distribution in the foreground
cluster at $z=0.33$, which is essential to reproduce the non-concentric
configuration of the two arcs.  The best-fit parameters for the
galaxy-scale mass component are in good agreement with the observed
light distribution and velocity dispersion of G1 -- the mass is
centered on G1 with an ellipticity of $\epsilon = 0.31\pm 0.01$ at a
P.A.\ of $103\pm 2$\,deg and a velocity dispersion $\sigma =230\pm
5$\,km\,s$^{-1}$.  The cluster contributes an additional shear of
$\gamma=0.15$ along a direction tangential to the cluster center, which
produces the offset in the centers of curvature of N and S.  This lens
model suggests a combined magnification for the arcs of $28\pm 3$.  The
model forms the arcs N/S from lensing of a single, extended
background source with a scale size of $\sim 1$\,kpc.  However, to
reproduce the faint western extension of N we require a second
background source, offset from the primary source by $\sim 0.3''$ ($\sim
2$\,kpc).  The proximity of these two sources means that they likely represent
two star forming knots within a single galaxy.  In this regard, we note
that if the galaxy wasn't lensed, it would appear as a single elongated
source even in deep ACS imaging.  A more detailed description of the
mass model for the lensing galaxy is given in Dye et al.\ (2007).

\subsection{Intrinsic Properties}

Having determined the magnification of the system we can estimate the
intrinsic luminosity of the background galaxy. Firstly we note
that the characteristic apparent magnitude of $z\sim 3$ LBGs from
Shapley et al.\ (2001) is $K^\ast = 22.52\pm 0.25$. Hence the combined
$K'$-band magnitude of the arcs N/S corresponds to an apparent
luminosity of $28 L^\ast_V$!  Correcting for the lensing magnification
of $\sim 28\pm 3$, the arcs' intrinsic apparent magnitude translates to
$K\sim 22.6\pm 0.2$, consistent with $L_V^\ast$ for the $z=3$ LBG
population (c.f.\ Allam et al.\ 2007). Similarly, comparing the colors
of the arcs, $(R_{606}-K')\sim 1.6$, with typical $z\sim 3$ LBGs from
Shapley et al.\ (2001), which have $(R-K)\sim 1.0\pm 0.6$, we find that
the arcs are slightly redder than the typical LBG but are within the
1-$\sigma$ scatter for the population.

Finally, to estimate the star formation rate in the galaxy we use the
$R$-band continuum magnitudes of the arcs to determine a restframe
1500\AA\ luminosity of $L_{1500}\sim
4.6\times10^{30}$\,ergs\,s$^{-1}$\,Hz$^{-1}$, which translates into a
star formation rate of 640\,M$_\odot$\,yr$^{-1}$, adopting a Salpeter
IMF with an upper-mass cut-off of 100\,M$_\odot$ (Kennicutt 1998) and
without any correction for reddening or magnification.  Correcting for
the estimated magnification and UV reddening we derive an intrinsic
star formation rate of $\sim 100$\,M$_\odot$\,yr$^{-1}$.

\section{Conclusions}

We report the discovery of a LBG at $z=3.0743$ which is seen as a pair
of arcs with an apparent magnitude of $r=20.3$, equivalent to
$28L^\ast_V$.  This very bright apparent magnitude results from
gravitational lensing by a galaxy at $z=0.73$ and a cluster
at $z=0.33$ which provide a combined magnification of $28\pm 3\times$.
Correcting for this magnification we show that LBG\,J213512.73$-$010143
(the ``Cosmic Eye'') is a compact $\sim L^\ast_V$ LBG with moderate
dust reddening, $E(B-V)\sim 0.4$, and a star formation rate of $\sim
100$\,M$_\odot$\,yr$^{-1}$.  We illustrate the wide range of spectral
features which can be identified in a modest signal-to-noise,
moderate-resolution spectrum of the galaxy and look forward to future
observations at higher signal-to-noise and resolution.  The importance
of this system is the opportunity it provides to obtain restframe UV,
mid-infrared and millimeter wavebands observations of a typical $z\sim
3$ LBG at the resolution and signal-to-noise which will only become
available for unlensed examples with the commissioning of ELTs, the
{\it James Webb Space Telescope (JWST)} or the Atacama Large Millimeter
Array (ALMA) respectively.  Such observations will provide unique
information on the elemental abundances, star formation, gas mass and
dynamics of this young galaxy, facilitating a range of studies
including the calibration of the various indicators proposed to trace
the metallicity of gas in distant galaxies.  This system can thus act
as a path-finder for the science which will be done with ELTs, JWST and
ALMA when they are completed.

\acknowledgments
We thank Max Pettini for help and useful discussions and an anonymous
referee for constructive comments which improved this work.  IRS and
GPS acknowledge support from the Royal Society, AMS acknowledges
support from PPARC.

\end{document}